\documentclass[twocolumn]{aastex62}
\usepackage{color,soul}

\newcommand{\SNeSampleSize}{160}
\newcommand{\SNeSpecSize}{1551}

\graphicspath{{./}{figures/}}

\shorttitle{Optimal Classification of SESNe}
\shortauthors{Williamson et al.}

\begin{document}

\title{Optimal Classification and Outlier Detection for Stripped-Envelope Core-Collapse Supernovae}

\author{Marc Williamson}
\affil{New York University}

\author{Maryam Modjaz}
\affiliation{New York University}
\affil{Center for Computational Astrophysics, Flatiron Institute,162 5th Avenue, 10010, New York, NY, USA}

\author{Federica B. Bianco}
\affil{Center for Computational Astrophysics, Flatiron Institute,162 5th Avenue, 10010, New York, NY, USA}
\affiliation{University of Delaware Department of Physics and Astronomy}
\affiliation{University of Delaware Joseph R. Biden Jr. School for Public Policy and Administration}
\affiliation{University of Delaware Data Science Institute}
\affiliation{New York University Center for Urban Science and Progress}

\begin{abstract}
In the current era of time-domain astronomy, it is increasingly important to have rigorous, data driven models
for classifying transients, including supernovae. We present the first application of Principal Component
Analysis to the photospheric spectra of stripped-envelope core-collapse supernovae. We use one of the largest compiled optical datasets of stripped-envelope supernovae, containing $\SNeSampleSize$ SNe and $\SNeSpecSize$ spectra. We find that the first 5 principal components capture 79\% of the variance of our spectral sample, which contains the main families of stripped supernovae: Ib, IIb, Ic and broad-lined Ic. We develop a quantitative, data-driven classification method using a support vector machine, and explore stripped-envelope supernovae classification as a function of phase relative to V-band maximum light. Our classification method naturally identifies ``transition'' supernovae and supernovae with contested labels, which we discuss in detail. We find that the stripped-envelope supernovae types are most distinguishable in the later phase ranges of $10\pm5$ days and $15\pm5$ days relative to V-band maximum, and we discuss the implications of our findings for current and future surveys such as ZTF and LSST. 

\end{abstract}



\section{Introduction} \label{sec:intro}

Supernova classification is a longstanding challenge in the astronomical community. The first spectral classification of supernovae (SNe) was introduced by Minkowski (\citeyear{minkowski1941spectra}) who defined two classes, Type I (Hydrogen absent) versus Type II (Hydrogen present). This broad criterion is still in use today, and multiple subclasses were added as the number of SNe spectra increased and spectral differences were observed (for a comprehensive review of SNe classifcation see Filippenko  \citeyear{filippenko1997optical}; Gal-Yam \citeyear{gal2017classification_rev}). In this work, we focus on stripped-envelope core-collapse supernovae (SESNe; Clocchiatti et al. \citeyear{clocchiatti1997sn}), which are the deaths of massive ($>8M_{\odot}$) stars that have lost part or all of their outer Hydrogen and Helium layers. The diversity of the amount of these elements remaining in the outer envelopes of the stellar progenitors at the time of explosion is the likely explanation for  the classification into three major SNe classes: Type Ib (spectra have conspicuous He features), Type IIb (spectra show strong H at early phases, He features at later phases), and Type Ic (no prominent H nor He features in spectra). For more detailed review of SESNe see \cite{filippenko1993type,matheson2001optical,woosley2002evolution,modjaz2014optical,liu16}. Over the last 20 years, the class of broad-lined SNe Ic (Ic-bl) has emerged with members showing spectra devoid of strong lines of H and He, but with broad lines that indicate expansion velocities between 15000 and 20000 km/s \citep{modjaz_icbl,prentice17,sahu2018broad}. In addition, the Ic-bl type is the only SN type associated with long-duration gamma ray bursts \citep[for reviews see][]{woosley2006sn_grb,modjaz2011grb,cano2017observer}.

\begin{deluxetable*}{CCCCC}[ht!]
\tablecaption{SESNe Dataset (includes compilations from L\&M14\tablenotemark{a}, L16\tablenotemark{a}, M16\tablenotemark{a}, and new additions below)\label{tab:datatable}}
\tablecolumns{5}
\tablenum{1}
\tablewidth{0pt}
\tablehead{
\colhead{Phase\tablenotemark{a}} &
\colhead{Ib} &
\colhead{IIb} &
\colhead{Ic} & 
\colhead{Ic-bl} \\
 & 
\colhead{$\left(N_{\textnormal{SNe}}, N_{\textnormal{Spec}}\right)$} &
\colhead{$\left(N_{\textnormal{SNe}}, N_{\textnormal{Spec}}\right)$} & 
\colhead{$\left(N_{\textnormal{SNe}}, N_{\textnormal{Spec}}\right)$} & 
\colhead{$\left(N_{\textnormal{SNe}}, N_{\textnormal{Spec}}\right)$}}
\startdata
0\pm5 & (28,81) & (21,62) & (27,79) & (17,74) \\
5\pm5 & (22,68) & (19,54) & (21,61) & (17,59) \\
10\pm5 & (23,54) & (18,41) & (21,47) & (15,36) \\
15\pm5 & (19,44) & (17,35) & (18,40) & (13,30) \\
        &        &          &        &         \\ \hline\hline
        &  &    \textbf{New SNe (Added to Liu et al Sample)}&       &        \\
\textnormal{SN Name}& \textnormal{SN Type}&\textnormal{Phases\tablenotemark{a}}&&\textnormal{Ref\tablenotemark{b}} \\\hline
\textnormal{SN2010as}& \textnormal{IIb}&\textnormal{-14,-13,-12,-11,-10,-9,-9,-9,-9,-8,-8,-8,-6,-6,-6,-5,-5,-5,6,19,(+6)}& &\textnormal{F14}\\
\textnormal{SN2011hs}& \textnormal{IIb}&\textnormal{-9,-8,-8,-7,-6,-5,-5,-5,4,5,8,25,25,26,27,28,28,53,58,(+4)}& &\textnormal{B14}\\
\textnormal{SN2012au\tablenotemark{c}}& \textnormal{Ib}&\textnormal{-6,-1,10,21,33,48,57,67,73,(+2)}& &\textnormal{T13}\\
\textnormal{SN2012P}& \textnormal{IIb}&\textnormal{-11,-8,-7,-2,1,8,26,29,31,(+1)}& &\textnormal{F12}\\
\textnormal{SN2013df}& \textnormal{IIb}&\textnormal{-14,-11,-4,-4,-4,0,4,11}& &\textnormal{S16,C16}\\
\textnormal{SN2013ge}& \textnormal{Ic}&\textnormal{-15,-14,-13,-4,-3,5,6,12,13,15,16,19,30,33,34,}& &\textnormal{D16}\\
& & \textnormal{37,38,40,43,44,46,46,62,65,67,70,70,71,(+9)}& &\\
\textnormal{SN2014ad}& \textnormal{Ic-bl}&\textnormal{27,27,37,37}& &\textnormal{YG12}\\
\textnormal{LSQ14efd}& \textnormal{Ic}&\textnormal{-12,-11,-4,-4,-4,4,4,17,17,17,23,23,23,32,32,32,32}& &\textnormal{S15}\\
\textnormal{iPTF15dtg}& \textnormal{Ic}&\textnormal{-16,-2,6,15,45,64,78,(+1)}& &\textnormal{T16}\\
\textnormal{SN2016coi}&\textnormal{Ic-bl}&\textnormal{-13,-10,-8,-2,-1,0,2,6,7,10,23,24,30,42,62,85,(+3)}&&\textnormal{P18}\\
\textnormal{SN2016gkg}&\textnormal{IIb}&\textnormal{-18,-18,-18,-17,-16,-15,-14,-11,-10,-7,-5,-1,-1,0,1,15}&&\textnormal{T17}\\
\textnormal{SN2017ein}& \textnormal{Ic}&\textnormal{-7,12,15,18,22,38,47,51,53}& &\textnormal{VD18}
\enddata

\tablenotetext{a}{Phases are rounded to the nearest integer and are in rest frame, relative to the date of V-band maximum. The number of nebular phase spectra (phase $>$90 days) are included in parentheses, but they are not used in our analysis.}

\tablenotetext{b}{References: L\&M14--Liu\&Modjaz (\citeyear{liu2014supernova}), L16--Liu et al (\citeyear{liu16}), M16--Modjaz et al (\citeyear{modjaz_icbl}), F14--Folatelli et al (\citeyear{folatelli2014SN10as}), B14--Bufano et al (\citeyear{bufano2014SN11hs}), T13--Takaki et al. (\citeyear{takaki2013luminous}), F12--Fremling et al. (\citeyear{fremling2016ptf12os}), S16--Szalai et al. (\citeyear{szalai2016continuing}), C16--Childress et al. (\citeyear{childress2016anu}), D16--Drout et al (\citeyear{drout2016SN13ge}), YG12--Yaron, Galyam (\citeyear{yaron2012wiserep}), S15--Smartt et al. (\citeyear{smartt_pessto}), T16--Taddia et al (\citeyear{taddia2016iptf15dtg}), P18--Prentice et al (\citeyear{prentice2018SN16coi}), T17--Tartaglia et al. (\citeyear{tartaglia2017progenitor}), VD18--Van Dyk et al (\citeyear{van2018SN17ein}).}

\tablenotetext{c}{The spectra of SN2012au could not be included in our PCA analysis because they were too noisy.}
\end{deluxetable*}
Current SESNe classification methods can be grouped broadly into two categories: template matching and specific feature techniques. The most used template matching algorithms are the Supernova Identification code (SNID) \citep{snid} and Superfit \citep{howell2005gemini}. These codes match new spectra to a library of previously classified supernovae using cross-correlation and chi-squared statistics respectively, yielding a quantitative measure of similarity between spectra of previously known SNe and the spectrum of a new transient. By incorporating more than just the best match into a classification scheme (e.g. Quimby et al. \citeyear{quimby2018spectra}), template matching can distinguish the major SESNe classes. However, template matching has some downsides. It is difficult to gain physical insight into stellar progenitors from a simple similarity measure. In addition, template matching classification methods do not directly yield a physical understanding of the differences between different classes. The second category of classification techniques focuses on characterizing specific spectral features (i.e. line depth or width and line intensity or velocity) at particular wavelengths \citep{sun17,prentice17}. These specific feature techniques allow for more physical interpretation than template matching, but they do not use all of the information available in a spectrum.

In this paper, we propose a new classification technique for SESNe using Principal Component Analysis (PCA) combined with Support Vector Machine (SVM). PCA is a dimensionality reduction algorithm that linearly transforms data in order to capture as much information as possible in the smallest number of transformed features, called principal components (PC's). PCA has been previously applied to attempt to understand the diversity of SNe Ia subtypes \citep{cormier2011study,sasdelli2014metric} and nebular phase superluminous supernovae \citep{nicholl2019nebular}, but this is the first application of PCA to SESNe in the photospheric phase. After applying a PCA decomposition to our SESNe spectral dataset, we use a multi-class linear SVM, a supervised learning method, to classify our SNe. This work is the first application of such machine learning techniques to spectroscopically classifying SESNe.

Our PCA and SVM based algorithm allows continuous, quantitative classification that reflects the physical properties of SESNe stellar progenitors. Instead of the traditional SN classification with four discreet classes (IIb, Ib, Ic, Ic-bl), our classification method facilitates better understanding of which SNe are representative of their class, and which are ``transition'' objects, and comparison between the SESNe mean spectra and our constructed eigenspectra allows us to physically interpret our results. New and upcoming data releases by the Berkeley group \citep{shivvers2018berkeley} and the Palomar Transient Factory \citep[PTF;][]{fremling_ptf_sesne,taddia_ptf_icbl}, and new transient observing projects like the Zwicky Transient Factory \citep[ZTF;][]{bellm2014zwicky} and the Large Synoptic Survey Telescope \citep[LSST;][]{ivezic2008lsst} will drastically increase the number of SESNe spectra. In this new data-rich context, a continuous, and data-driven classifier will be crucial for addressing some of the most interesting outstanding questions pertaining to SESNe.

\section{SESNe Spectral Dataset} \label{sec:data}
In this section, we describe the spectral dataset used in this work and the preprocessing applied to the data before our analysis is performed. We expand the SESNe spectral library produced and compiled in Modjaz et al. (\citeyear{modjaz2014optical}), Liu \& Modjaz (\citeyear{liu2014supernova}), Liu et al. (\citeyear{liu16}) and Modjaz et al. (\citeyear{modjaz_icbl}) to include available spectra from SNe published through August 2018. We use the same criteria for inclusion of new data as Liu et al. (\citeyear{liu16}): well-typed SNe with light curves for which a date of maximum can be extracted. The dataset contains $\SNeSampleSize$ SNe and $\SNeSpecSize$ spectra. We exclude SNe Ib-n, SNe Ib-Ca, superluminous supernovae, and SNe that transition between normal and excluded types. We restrict the spectra in our sample to the optical wavelength range 4000$\AA\ $to 7000$\AA\ $ since the vast majority of our SNe have observed fluxes in this wavelength range, and this range contains features of both H and He that drive the classification. For newly added SN spectra obtained from the literature or directly from authors, we follow the same preprocessing steps detailed in Liu et al. (\citeyear{liu16}) that were used in subsequent papers of our group \citep{liu16,modjaz_icbl}. The preprocessing is briefly summarized as follows: when newly added spectra lack a date of V-band maximum (but do have a date of maximum in other bands), we convert their date of maximum to V-band using the process described by Bianco et al (\citeyear{bianco_vband_conv}). Spectra are redshift corrected when necessary, and the continuum removal and normalization (spectra are scaled by their means to have relative fluxes) is performed with tools within the SNID framework \citep{snid}. In the few cases where telluric lines are present in the spectra, the tellurics are removed using linear interpolation consistent with the procedure in Liu et al (\citeyear{liu16}). Small gaps in the spectra are similarly interpolated before a fourier based smoothing is applied \citep{liu16}. The bandpass filter used by SNID for classification purposes is not applied. A summary of our dataset can be found in Table \ref{tab:datatable}, and the SNID templates of the newly added SNe are released on our github page\footnote{\url{https://github.com/nyusngroup/SESNtemple/tree/master/SNIDtemplates}}.

\begin{figure}[ht!]
\includegraphics[width=\columnwidth]{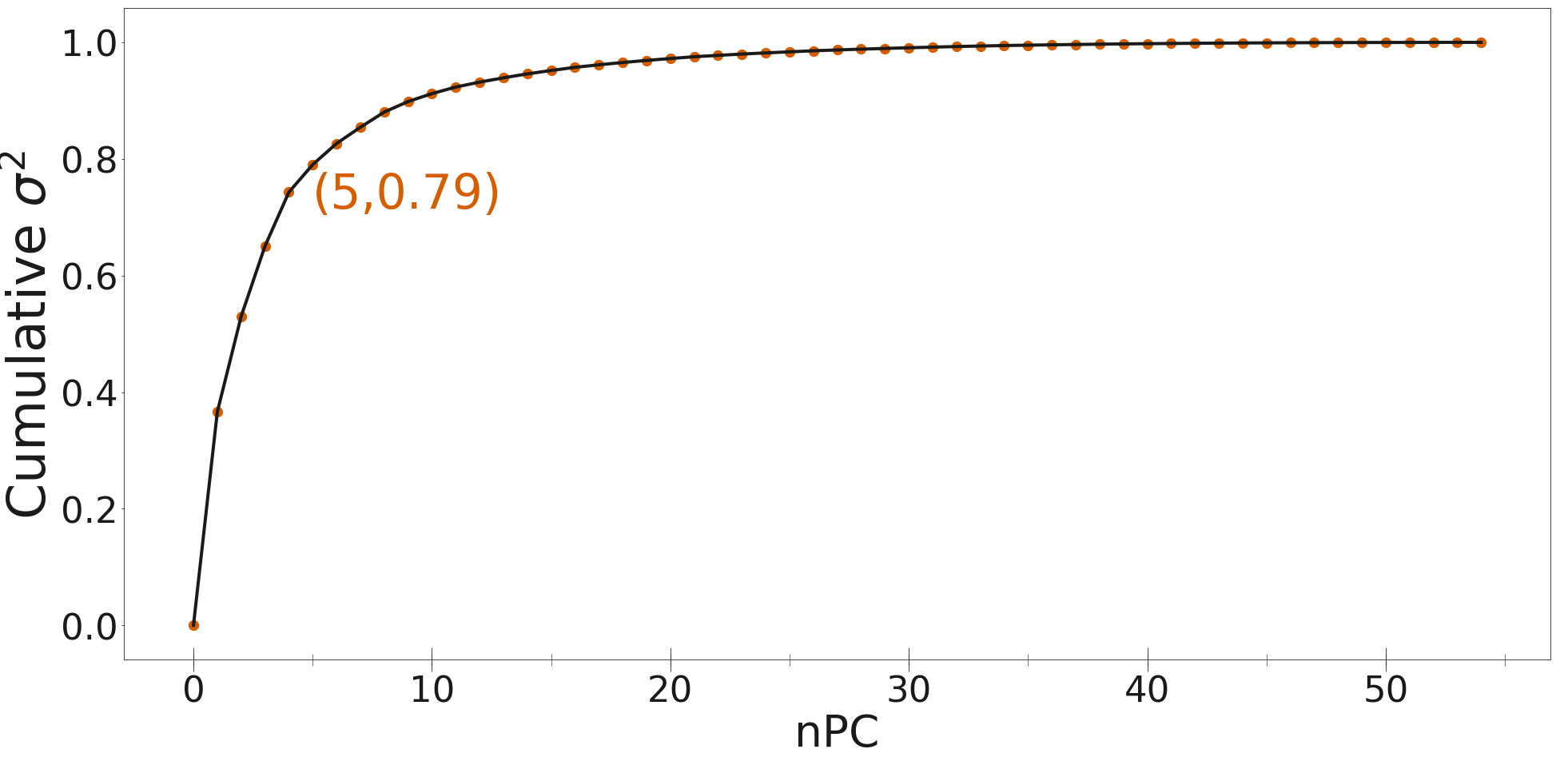}
\caption{Cumulative fraction of variance of the entire SESNe dataset captured by nPC eigenspectra. The first 5 eigenspectra capture 79\% of the sample variance. \label{fig:pcacum}}
\end{figure}

\section{Methods} \label{sec:meth}
In this section, we present a brief background on the two machine learning methods used in our analysis, PCA and SVM, as well as details on our specific application. For both methods, we use the \texttt{scikit-learn}\footnote{\url{https://scikit-learn.org/stable/index.html}} implementation \citep{pedregosa2011scikit}. For a detailed review of PCA theory see Pearson (\citeyear{pearson1901liii}) and Jolliffe (\citeyear{jolliffe2011principal}), and for a detailed review of SVM theory see Vapnik (\citeyear{vapnik1998support}). Our research is reproducible: all code and raw data is accessible on github \footnote{\url{https://github.com/nyusngroup/SESNspectraPCA}}.

\subsection{PCA--Derivation of Eigenspectra}\label{sec:pca}
\begin{figure}[ht!]
\includegraphics[width=\columnwidth]{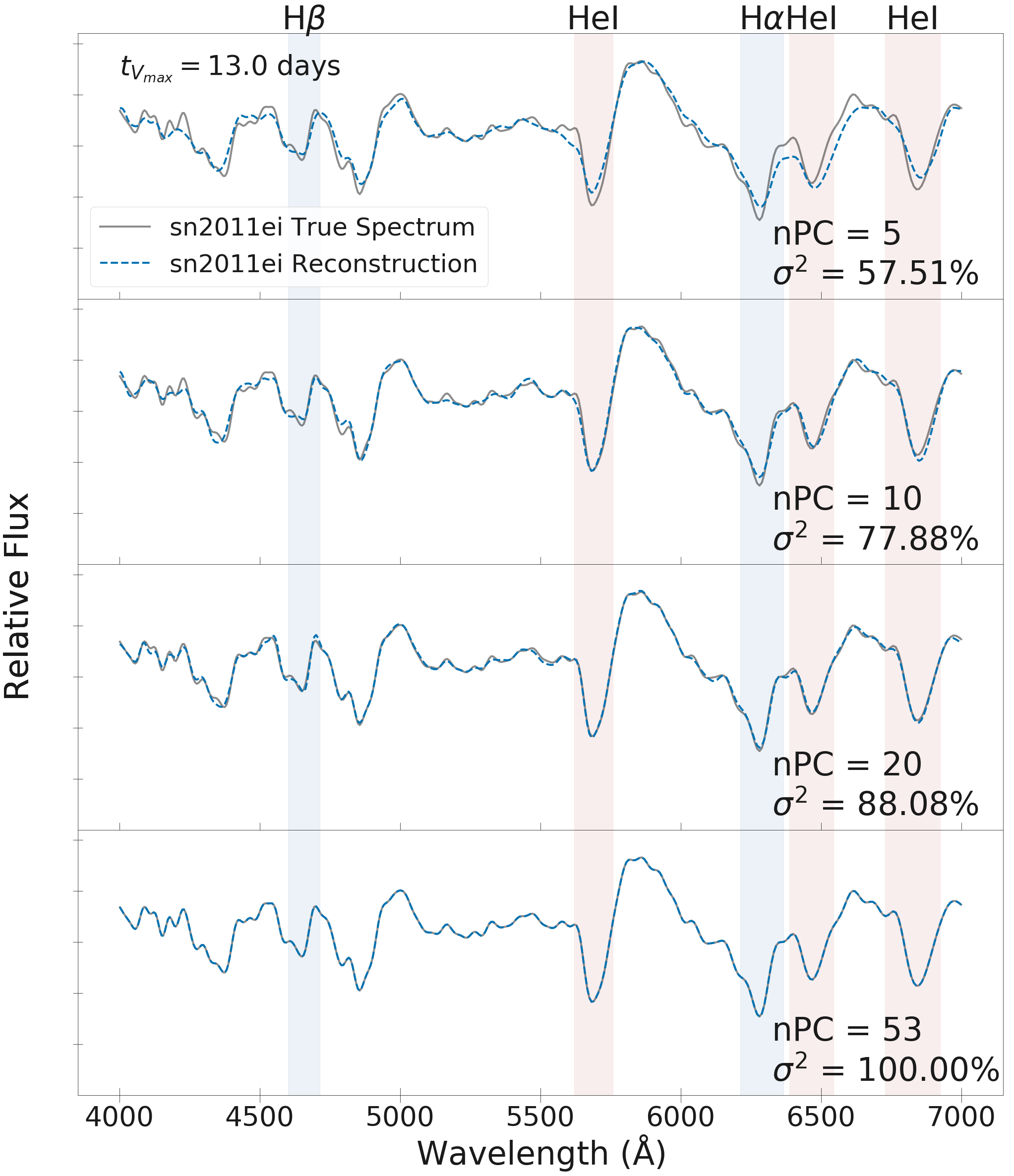}
\caption{Reconstructions of the spectrum of SN type IIb SN2011ei \citep{milisavljevic2013multi} at phase $t_{V_{max}}=13$ days. An increasing number of eigenspectra (nPC) is used to reconstruct the original spectrum from top to bottom.  As nPC increases, more features are captured, but 5 eigenspectra already capture the H and He features (indicated by shaded regions).\label{fig:reconstruct}}
\end{figure}

PCA is a dimensionality reduction technique based on singular value decomposition of a data matrix. The principal components are the eigenvectors of the covariance matrix of the data, and are therefore orthonormal. Each PC is a linear combination of the original data features (normalized fluxes) and therefore has the same wavelength range as our original data. We therefore use the term ``eigenspectra'' to describe the PC's and discuss their physical interpretation in Section \ref{sec:eig}. The eigenspectra are ordered according to how much variance from the mean of the dataset each component captures. Thus, the original spectra can be projected onto a subset of the eigenspectra while maximizing the amount of information retained. Figure \ref{fig:pcacum} shows the cumulative amount of variance of the entire dataset captured as a function of the number of PC's. The first five eigenspectra contain seventy-nine percent of the variance. Figure \ref{fig:reconstruct} shows an example supernova, SN2011ei (type IIb), reconstructed using increasingly larger numbers of eigenspectra. In the top panel, only the first five eigenspectra are used, and the large scale spectral features are  almost entirely reconstructed. For the purpose of classification, we mostly care about the large scale features, so considering only the first 5 eigenspectra of our PCA decomposition is a good first step to reduce the complexity of the problem. 

Since SNe change over time, in this work we apply a PCA decomposition to four different phase ranges of spectra: $0\pm 5$, $5\pm 5$, $10\pm 5$, and $15\pm 5$ days relative to V-band maximum. We present and discuss in detail the eigenspectra for the phase range $t_{V_{max}}=15\pm5$ days in Section \ref{sec:eig_mean_temp}. The time dependence of the eigenspectra as a function of phase is discussed in Section \ref{sec:eig_time}, but in general we find that there is very little change in the large scale features of a given eigenspectrum over time.

\subsection{SVM--A New Approach to SESNe Classification}\label{sec:svm}

For each of the four phase ranges in this work, we train a multi-class linear SVM without class weights and using the L2 (i.e. ``squared hinge'') loss function on the 2D projection of SESNe spectra onto each pair of the first 10 eigenspectra in order to understand which eigenspectra are most useful for classification. Specifically, we use the LinearSVC class from \texttt{scikit-learn} which implements SVM classification using LIBLINEAR \citep{fan2008liblinear} and employs the ``one-vs-rest'' approach to multi-class labeling and the ``winner-take-all'' approach to multi-class predictions: a binary linear SVM is trained to distinguish each class of SESNe from the rest of the population, and these binary classifiers are combined to make final decisions on predicting the labels of new data. Each binary SVM determines the optimal hyperplane that separates one class from the rest of the data. For each 2D projection, we randomly generate multiple train-test splits of the data (a random subset of 70\% of the data is used to train the SVM, while the remaining 30\% is used to test the ability of the SVM to accurately predict SNe classes). Using multiple train-test splits on each 2D projection allows us to report a mean test score for the SVM and to gain insight into the uncertainty of the SVM linear decision boundaries. The results of our SVM classification are discussed in Section \ref{sec:classify}.

\begin{figure*}[ht!]
\includegraphics[width=\textwidth]{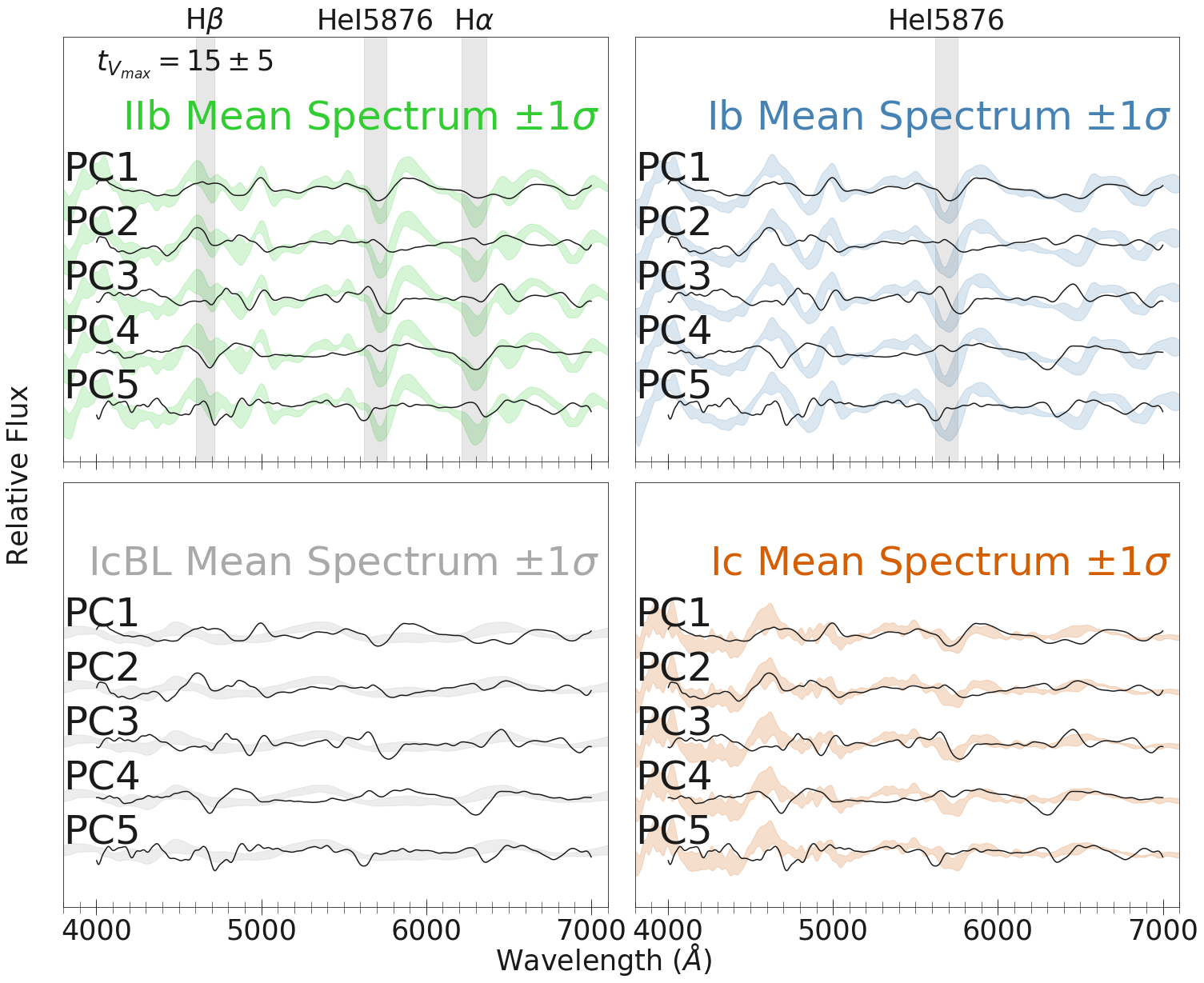}
\caption{Comparison of the first five eigenspectra at phase $t_{V_{max}}=15\pm5$ days, constructed using data of all SESNe types, with the mean spectra \citep{liu16,modjaz_icbl} for type SNe IIb (upper left), Ib (upper right), Ic-bl (lower left), and Ic (lower right). The eigenspectra are scaled by a factor of 2 and sign choice is made to facilitate comparing the relative structure of the principal components vs the mean spectra. PC1 and PC3 have a strong trough that lines up with the HeI5876 absorption feature in types IIb and Ib SNe. PC4 has strong troughs that line up with the H$\alpha$ and H$\beta$ absorption features in the type IIb mean spectrum.}\label{fig:pcatemplate}
\end{figure*}

\section{Physical Interpretations of Eigenspectra} \label{sec:eig}
One of the major benefits of our PCA and SVM based classification method is that we can physically interpret the eigenspectra using mean spectra of each of the SESNe classes. This allows us to understand why the SVM identifies certain eigenspectra as better classifiers than others, and how this behavior changes as a function of phase.

\subsection{Comparing Eigenspectra to SESNe Mean Spectra}\label{sec:eig_mean_temp}
The first few eigenspectra are the most important building blocks for reconstructing a spectrum from our dataset. Therefore, in order to understand any strong eigenspectra features, we compare the first five eigenspectra for the phase range $t_{V_{max}}=15\pm5$ days to the mean spectra for each of the four SESNe types, presented in Liu et al. (\citeyear{liu16}) and Modjaz et al. (\citeyear{modjaz_icbl}). The first 5 eigenspectra are plotted in Figure \ref{fig:pcatemplate}, along with the mean spectra for the four SESN subtypes. The principal components are naturally normalized, and we choose the sign of each component to properly represent the absorption features they capture. We highlight a few important features of each of the first 5 eigenspectra:

\begin{enumerate}
\item PC1 has a strong trough that lines up with the HeI5876 absorption feature present in both types IIb and Ib mean spectra, as well as the absorption feature in the Ic mean spectrum, (the cause of which is debated; Dessart \& Hillier \citeyear{dessart2010supernova}).
\item PC2 matches the Ic mean spectra closely up to $\lambda \approx 5500\AA\ $.
\item PC3 has small troughs in the H$\alpha$ and H$\beta$ regions in addition to a stronger trough in the HeI5876 region.
\item PC4 has strong troughs in the H$\alpha$ and H$\beta$ regions, but lacks a strong HeI5876 feature.
\item Due to the broadening of their features, SNe Ic-bl are effectively nearly featureless spectra, which result in a much smoother average spectrum than any of the first 5 PC's.
\end{enumerate}

These similarities between the eigenspectra and the SESNe mean spectra provide an excellent context to interpret the SVM classification. From Figure \ref{fig:pcatemplate}, we see that all of the SESNe types except Ic-bl have an absorption feature near $\lambda\approx5876\AA\ $ (although this feature is most likely not due to Helium for the Ic type). Moreover, as shown in Liu et al. (\citeyear{liu16}), this feature exists in the IIb, Ib, and Ic mean spectra even at early phases. Therefore we conclude that PCA generates eigenspectra that match previously identified important SESNe spectral features.

\begin{figure}[ht!]
\includegraphics[width=\columnwidth]{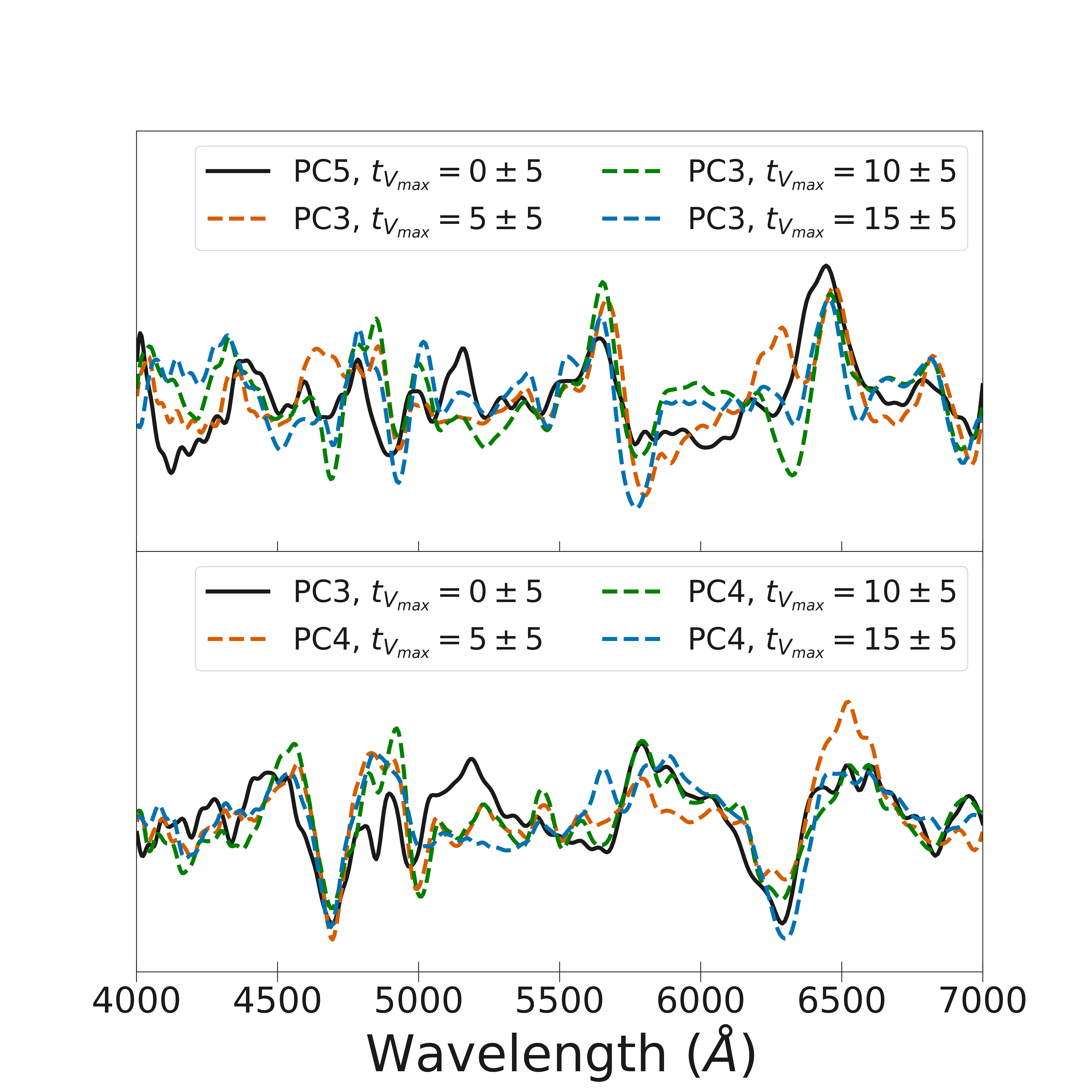}
\caption{Change in eigenspectrum order between $t_{V_{max}}=0\pm5$ days vs later phase ranges. PC5 at early times is equivalent to PC3 at later times as they capture the same features: H$\alpha$, H$\beta$, and HeI5876 as discussed in Section \ref{sec:eig_mean_temp}. Similarly, PC3 at early times is equivalent to PC4 at later times because they both primarily capture H$\alpha$ and H$\beta$ absorption. Otherwise, the important large scale features do not change with time.\label{fig:eig_time_fig}}
\end{figure}

\begin{figure*}
\includegraphics[width=\textwidth]{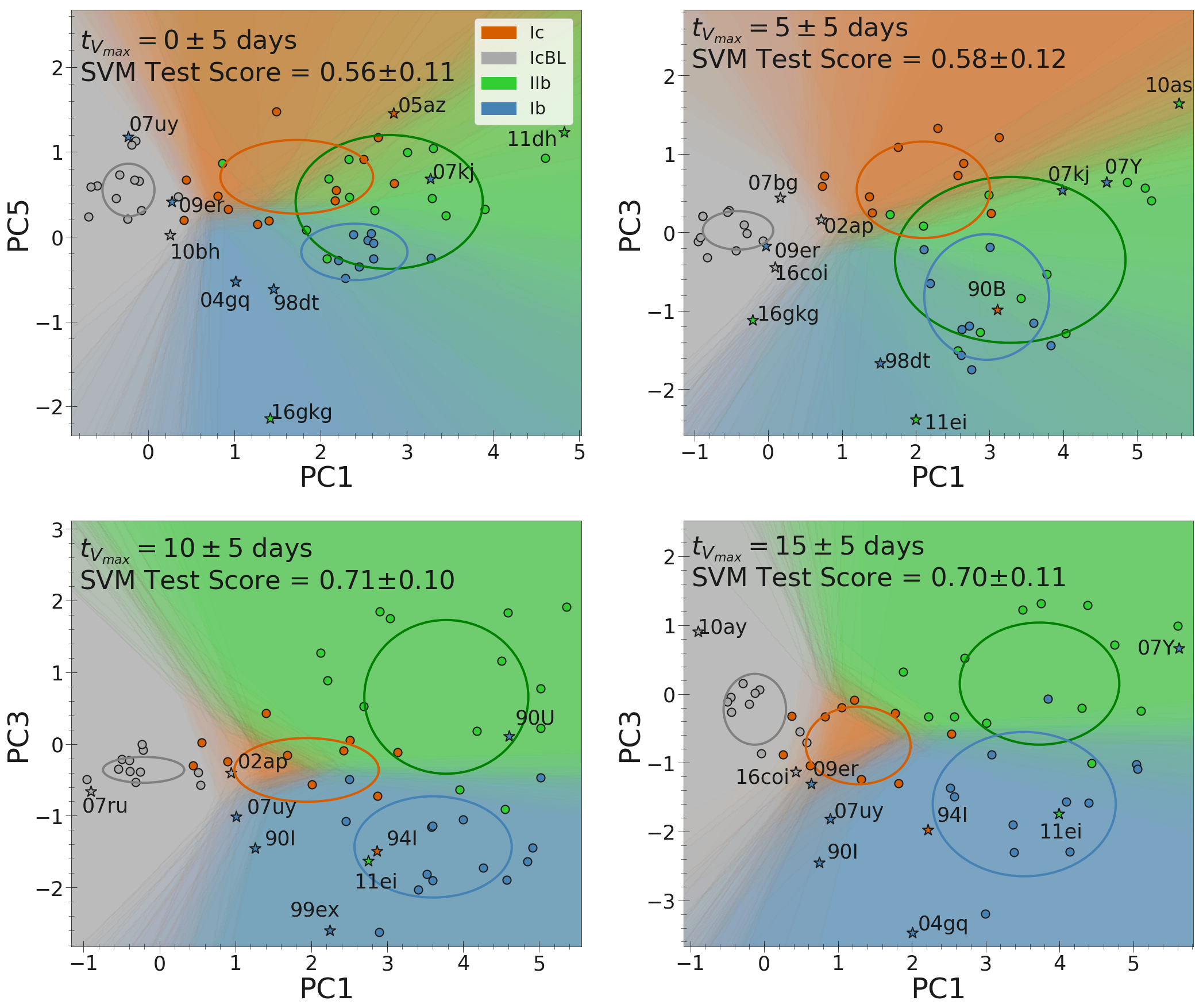}
\caption{Each panel shows the SESNe classification regions and linear decision boundaries for each SVM train-test split of the data. Ellipses represent the 1 standard deviation contour of the PC coefficients for each SESN type (excluding the peculiar SNe SN2007uy, SN2009er, and SN2005ek). Outliers of more than 2 standard deviations from the mean are marked with stars. The phase range $t_{V_{max}}$ is labeled in the upper left of each panel, along with the mean SVM test-score. PC1 vs PC3 provides the highest SVM test-score for each phase range except $t_{V_{max}}=0\pm5$ where PC1 vs PC2 has a slightly ($<1\sigma$) higher SVM test-score but very similar 1-$\sigma$ contour and SVM region overlap. \textit{Upper Left:} ($t_{V_{max}}=0\pm5$ days) There is large overlap between the IIb (green), Ib (blue), and Ic (orange) 1-$\sigma$ contours, and between the SVM IIb and Ib region, and the IIb and Ic region (as indicated by the region boundaries changing significantly for different train-test splits and the colors bleeding into each other). \textit{Upper Right:} ($t_{V_{max}}=5\pm5$ days) As as $t_{V_{max}}=0\pm5$ days, there is overlap between the IIb, Ib, and Ic 1-$\sigma$ contours, and the corresponding SVM regions. \textit{Lower Left:} ($t_{V_{max}}=10\pm5$ days) The IIb, Ib, Ic, and Ic-bl 1-$\sigma$ contours are completely separated. Each colored SVM region is well defined and stable for different train-test splits, and the SVM test score is highest. The Ic (orange) SVM region has collapsed and the IIb (green) SVM region has expanded. \textit{Lower Right:} ($t_{V_{max}}=15\pm5$ days) SESNe type 1-$\sigma$ contours are well-separated and the SVM regions are stable.}\label{fig:marg}
\end{figure*}

\subsection{Time Evolution of Eigenspectra}\label{sec:eig_time}
In Section \ref{sec:eig_mean_temp} we present the eigenspectra only for the phase range $t_{V_{max}}=15\pm5$ days because we find the SESNe types to be maximally separated at this phase, as we show in Section \ref{sec:classify}. Here we discuss how the eigenspectra change as a function of time. We have calculated and compared the first five eigenspectra for each of the phase ranges $0\pm5$ days, $5\pm5$ days, $10\pm5$ days, and $15\pm5$ days, relative to V-band date of max. We find that there is very little change for a given eigenspectrum across different phases. However, there is a slight change in the ordering of the first five eigenspectra between the later phase ranges and the $t_{V_{max}}=0\pm5$ day phase range. Figure \ref{fig:eig_time_fig} shows that PC5 at phase $t_{V_{max}}=0\pm5$ days corresponds (i.e. is most similar) to PC3 of the later phase ranges, and PC3 at phase $t_{V_{max}}=0\pm5$ corresponds to PC4 of the later phases. In the later phase ranges, PC3 is the eigenspectrum with weak troughs in the H$\alpha$ and H$\beta$ regions and a strong trough in the HeI5876 region. Thus it is not surprising that this eigenspectrum captures less variance of the sample in the earliest phase range. Liu et al. (\citeyear{liu16}) showed that the pseudo-equivalent line width (pEW) of HeI5876 in SNe types IIb and Ib are at their lowest values near V-band maximum and increase over time. PC4 in the later phases, which consists of two strong troughs at the H$\alpha$ and H$\beta$ wavelengths, is more highly ranked in the $t_{V_{max}}=0\pm5$ phase range because the H$\alpha$ absorption feature is very strong in type IIb spectra even at early phases.

\section{SVM Classification Results} \label{sec:classify}

Our goal is to create a method that reproduces the standard empirical classification scheme which classifies SNe spectra using the H and He features. We apply SVM to every 2D projection of the first 10 eigenspectra (following the procedure in Bianco et al. \citeyear{Bianco:2016:HIN:2993422.2993570}), and we find that the highest test scores (see Section \ref{sec:svm}) are always associated with a pair of the top 5 eigenspectra. These results are consistent with both Fig \ref{fig:pcacum} and Fig \ref{fig:reconstruct}, where the first 5 eigenspectra are sufficient to capture 79\% of the spectral variance in our sample and reproduce the spectrum of SN2011ei, respectively. In Figure \ref{fig:marg}, each panel corresponds to a different phase range, and in each panel we show the 2D plane which lead to the highest classification score. In the case where one phase range has multiple optimal planes (i.e. test scores are consistent within 1$\sigma$), we choose the eigenspectra pair to be physically consistent with PC1 vs PC3 at late times (i.e. $t_{V_{max}} =10, 15$ days) because this pair of eigenspectra produces the highest SVM test score across all phases and the least amount of overlap of the 1-$\sigma$ contours of the PCA coefficients for the different SESN classes. We find that we can recreate the SNID labels of our dataset. Furthermore, we find that the optimal phase ranges for classifying SESNe are $t_{V_{max}}=10\pm5$ days and $t_{V_{max}}=15\pm5$ days, as opposed to at maximum light ($t_{V_{max}}=0\pm5$ days). This is important in a future that, with the advent of LSST, will see an overwhelming number of SN discoveries, and a radical pressure on the urgency of spectroscopic follow-up for classification. Lowering the pressure on immediate follow-up for one type of transient (SESNe) alleviates pressure on the follow-up facilities altogether.

\subsection{Classification in the PC1 vs PC3 Projection}\label{sec:pc1vpc3}

Figure \ref{fig:marg} shows the two-dimensional projection of our SESNe spectra onto the optimal eigenspectra pairs that maximally separate subclasses: PC1 vs PC3 for $t_{V_{max}}=5\pm5,10\pm5,15\pm5$ days, and PC1 vs PC5 for $t_{V_{max}}=0\pm5$ days (as we described in Section \ref{sec:eig_time}, PC5 at $t_{V_{max}}=0\pm5$ corresponds to PC3 in the later phase ranges). The colored regions illustrate the linear SVM decision boundaries. Boundaries for 50 different 70\%-30\% train-test splits of the data are shown, thus assessing the statistical robustness of the decision boundaries. The SVM test-score, a measure of the accuracy of the classification, is indicated in each figure panel, including uncertainties generated from the 50 train-test splits. Colored ellipses in each panel represent the 1-standard-deviation (1-$\sigma$) contours of the PC coefficients for the different SESNe types. We have not included SNe Ib-pec (e.g. SN2007uy, SN2009er Modjaz et al. \citeyear{modjaz2014optical}) nor SNe Ic-pec (e.g. SN2005ek Drout et al. \citeyear{drout2013fast}) in the calculation of the ellipses (but we do show the datapoints of these peculier subtypes).

Both SVM test-score and the 1-$\sigma$ contours allow us to evaluate the success of our classifying scheme. The highest SVM test scores $(.71\pm.10\text{ and }.70\pm.11)$ are achieved at $t_{V_{max}}=10\pm5$ and $15\pm5$ days respectively. These scores however are statistically consistent at the 1 or 2 sigma level with the lower test scores of the earlier phases. Nonetheless, the SESN classes are more compactly clustered and separated at later times as shown by the 1-$\sigma$ contours that are maximally separated at these later phases. Therefore we find that the optimal time for classifying SESNe spectra is later than- ($t_{V_{max}}=10\pm5,15\pm5$ days) rather than at- or near-peak ($t_{V_{max}}=0\pm5,5\pm5$ days).

PC1 is a poor choice of eigenspectrum for SESNe classification at early times because the HeI5876 absorption feature in Ic, IIb, and Ib spectra has not had time to strengthen. In the phase ranges $t_{V_{max}}=10\pm5$ and $t_{V_{max}}=15\pm5$ days, PC1 and PC3 both become more effective at distinguishing between SESNe spectral types, with less overlap in the 1-$\sigma$ contours and a higher SVM test-score. In particular, we find that the PC1 coefficients of SNe types IIb and Ib increase (while SNe Ic PC1 coefficients remain relatively unchanged) as phase increases. Since PC1 captures the strong feature at $\lambda\approx5600-5800\AA\ $ which is due to He in SNe Ib and IIb, this behavior is consistent with Liu et al. (\citeyear{liu16}), which found that the pseudo-equivalent width (pEW) of the HeI absorption features in SNe types IIb and Ib increases as a function of phase. Figure \ref{fig:marg} also shows that as phase increases, PC3 becomes better at distinguishing between SNe types IIb (green region) and Ib (blue region). Specifically, the SNe type IIb PC3 coefficients systematically increase with increasing phase. Since PC3 captures the H$\alpha$ and H$\beta$ features, this behavior is consistent with the strengthening of the H$\beta$ absorption feature in SNe IIb mean spectra shown in Liu et al. (\citeyear{liu16}).

The SNe Ic-bl region (gray) is reasonably well separated from the other SESNe types at all phases in Figure \ref{fig:marg}. However, note that the Ic-bl data and the corresponding 1 standard deviation contour is centered near the origin in every panel. Moreover, we find the PC coefficients of the SNe Ic-bl to be clustered around zero in every two dimensional projection of the first five eigenspectra. This is expected because the SNe Ic-bl mean spectra do not have a strong absorption feature due to HeI5876, even if it were highly broadened \citep{modjaz_icbl}.

\subsection{Transition Supernovae and Type Outliers in PCA Space}\label{sec:transition}
One major benefit of our work is that the PC coefficients of the SESNe in our sample are continuous, and therefore well suited for capturing the physical continuity of chemical abundances in SNe ejecta. This behavior is particularly useful for objectively identifying ``transition'' SNe, which often have debated classification in the literature due to spectra that resemble more than one SESN type. Our method also identifies outliers in a particular class that are extreme versions of the SESN type, but not ``transition'' SNe. In Figure \ref{fig:marg} we label all SNe in each panel that are more than 2 standard deviation outliers and discuss them below.

\subsubsection{Type Ib Outliers}\label{sec:ib-bl}
Figure \ref{fig:marg} shows two SNe Ib that are consistently strong outliers: SN2007uy and SN2009er. These two supernovae either appear within the SNe Ic-bl (gray) region or close to the SVM decision boundary separating the Ic-bl and Ib regions (note that if SN2007uy or SN2009er does not appear in a panel of Figure \ref{fig:marg}, it is because we have no spectra in the corresponding phase range). SN2007uy and SN2009er have been previously identified in the literature \citep{modjaz2014optical} as peculiar members of the Ib class. Modjaz et al. showed that SN2007uy and SN2009er have broader features at higher velocities than normal SNe Ib spectra, in agreement with our results. We also find that SN1990I, SN1998dt, and SN2004gq are consistent outliers towards the Ic-bl region, although to lesser degrees than SN2007uy and SN2009er. Elmhamdi et al. (\citeyear{elmhamdi2004sn}) have previously identified SN1990I as having high velocity features atypical of a normal SN Ib, and Modjaz et al. (\citeyear{modjaz2014optical}) show that SN2004gq and SN1990I both have high absorption velocity He features compared to other SNe Ib spectra. The outliers SN1990I, SN1998dt, and SN2004gq may form a continuum of SN Ib spectra with higher than normal doppler shifts, while SN2007uy and SN2009er indicate the possibility for a continuum of SNe Ib spectra with varying amounts of line blending. SN1999ex was initially classified as an SN Ic, then changed to an SN Ib/c due to moderate HeI absorption features\citep{hamuy2002optical}. More recently, SN1999ex has been classified as an SN Ib \citep{modjaz2014optical}. We identify SN1999ex as an outlier in multiple 2D projections of PCA space, indicating that it is not a standard SN Ib nor a standard SN Ic.

We also identify SN1990U, SN2007kj, and SN2007Y as outliers in Figure \ref{fig:marg}. SN1990U (found in the green SN IIb region) has previously been considered as an SN Ic \citep{matheson2001optical} and more recently as an SN Ib \citep{modjaz2014optical}. Although we identify SN1990U as an outlier SN Ib in the PC1 vs PC3 projection, in the other projections it is a SN Ib, and in no projection is SN1990U located in the standard SN Ic region. Therefore our results support the reclassification of SN1990U as an SN Ib by Modjaz et al. (\citeyear{modjaz2014optical}). SN2007kj was previously classified as an SN Ib/c ``transition'' object \citep{leloudas2011properties} and more recently as a SN Ib \citep{modjaz2014optical}. We find that SN2007kj would be considered a strong outlier as an SN Ic in every 2D projection of the first 5 eigenspectra, while it is consistent with being a standard SN Ib in multiple 2D projections (not shown) other than PC1 vs PC3. Therefore we support the reclassification of SN2007kj as an SN Ib by Modjaz et al. (\citeyear{modjaz2014optical}). SN2007Y has been classified both as an SN IIb \citep{folatelli2014SN10as} and an SN Ib \citep{liu16}. In the PC1 vs PC3 2D projection, we find that SN2007Y falls in the IIb region, consistent with Folatelli et al. (\citeyear{folatelli2014SN10as}) who argued that SN2007Y is a SN IIb due to the strength and velocity of the HeI5876 feature. However, in another 2D projection (not shown), namely PC1 vs PC4 (strong H$\alpha$ and H$\beta$ features) at phases $t_{V_{max}}=5\pm5,15\pm5$ days, we find that SN2007Y falls in the Ib region, in agreement with Liu et al. (\citeyear{liu16}) who found that the H feature evolution of SN2007Y was consistent with SN Ib spectra. Thus our classification method captures the debate over the correct type for SN2007Y.

\subsubsection{Type IIb Outliers}
In Figure \ref{fig:marg} we label the following outlier SNe IIb: SN2010as, SN2011ei, and SN2016gkg. At early times, SN2010as appears on the decision boundary between types Ic (orange) and IIb (green), which is consistent with Folatelli et al. (\citeyear{folatelli2014SN10as}) who found that SN2010as exhibits weaker than normal He features at early times, in addition to weak H features. SN2011ei is a strong outlier in the PC1 vs PC3 2D projection. Milisavljevic et al. (\citeyear{milisavljevic2013multi}) showed that SN2011ei evolves quickly, losing its H features within a week after V-band maximum, to resemble a type Ib spectrum characterized by Helium features. Figure \ref{fig:marg} illustrates this evolution, with SN2011ei initially a standard IIb at phase $t_{V_{max}}=0\pm5$ days, then subsequently moving to the Ib region. However, Liu et al. (\citeyear{liu16}) showed that the H$\alpha$ equivalent width evolves differently for type IIb and Ib spectra (including SN2011ei), so SN2011ei is distinguishable as a SN IIb even at late times. When we consider the PC1 vs PC4 (strong H$\alpha$ feature) 2D projections (not shown) at the later phase ranges $t_{V_{max}}=5\pm5,10\pm5,15\pm5$ days, we find SN2011ei to be consistently within the IIb (green) region in agreement with Liu et al. (\citeyear{liu16}). SN2016gkg is classified as a SN IIb due to its H$\alpha$ absorption, but Tartaglia et al. (\citeyear{tartaglia2017progenitor}) showed that SN2016gkg exhibits stronger than normal Helium features even at early times, similar to an SN Ib. Figure \ref{fig:marg} captures this behavior, showing SN2016gkg as a strong outlier in the Ib (blue) region at $t_{V_{max}}=0\pm5$ days, but a more normal SN Ib in other 2D projections (not shown).

\subsubsection{Type Ic/Ic-bl Outliers}
We identify three Ic outliers, SN1990B, SN1994I, SN2005az, and six Ic-bl outliers, SN2002ap, SN2007bg, SN2007ru, SN2010ay, SN2010bh, and SN2016coi in Figure \ref{fig:marg}.SN1990B is currently considered an SN Ic, however it was initially classified as an SN Ib \citep{clocchiatti2001type}, which is consistent with our results in Figure \ref{fig:marg}. SN1994I is one of only a few SN Ic with many spectra taken over a range of wavelength regimes (e.g. \cite{filippenko1995type}, \cite{richmond1996ubvri}, \cite{immler1998evidence}), and it is considered a prototypical SN Ic. However, we find that SN1994I is considered an outlier in many 2D PCA projections, at multiple phases, as illustrated in Figure \ref{fig:marg}. Our results indicate that SN1994I may not be a prototypical SN Ic, confirming the spectroscopic analysis of Modjaz et al. (\citeyear{modjaz_icbl}) and the photometric analysis of Drout et al. (\citeyear{drout2011first}) and Bianco et al. (\citeyear{bianco_vband_conv}). SN2005az was initially classified as both an SN Ic \citep{aldering05azatel} and an SN Ib \citep{05azquimby}. Recently SN2005az has been classified as an SN Ic \citep{kelly2012core} using SNID based on the updated SESNe library from Modjaz et al. (\citeyear{modjaz2014optical}) and Liu et al. (\citeyear{liu16}). We find that SN2005az is inconsistent with being an SN Ib in the majority of 2D pojections, and when it is consistent with belonging to the Ib or IIb class, this is due to large overlap of the Ic and Ib/IIb regions at $t_{V_{max}}=0\pm5$ days. Meanwhile, there are some PCA 2D projections (PC2 vs PC4, not shown) where SN2005az is located within the SN Ic one standard deviation contour, so we support the classification of SN2005az as an SN Ic.

SN2002ap is claimed to be a relatively low energy SN Ic-bl compared to normal SNe Ic-bl events \citep{mazzali2002type} and has been classified as a normal SN Ic from radio observations \citep{berger2002radio}. We find that SN2002ap is indeed a potential transition object between the Ic and Ic-bl regions in Figure \ref{fig:marg}. Although SN2007bg is identified as an outlier at phase $t_{V_{max}}=5\pm5$ days, it is well within the Ic-bl SVM region (gray), and it no longer fulfills the outlier criterion at later phases, in agreement with the literature view that SN2007bg is a standard SN Ic-bl \citep{young2010two}. Similarly, although SN2007ru is marked as an outlier in the lower left panel of Figure \ref{fig:marg}, it is well within the SN Ic-bl SVM region and considered a standard SN Ic-bl \citep{sahu2009broad}. SN2010bh is considered a standard SN Ic-bl, although with slightly higher inferred explosion energy than other standard Ic-bl SNe \citep{chornock2010spectroscopic}. At late times (bottom right) of Figure \ref{fig:marg}, we find that SN2010ay is a strong SN Ic-bl outlier well within the Ic-bl region. SN2010ay is a particularly interesting Ic-bl because it has been proposed that SN2010ay was associated with an off-axis low luminosity gamma ray burst, due to its high absorption velocity, high peak luminosity, and low metallicity \citep{sanders2012sn}, combined with a lack of observed gamma rays. SN2016coi has broad spectral features in addition to a strong absorption feature generally attributed to HeI in the literature \citep{prentice2018SN16coi}, setting it apart from normal Ic-bl SNe. We find that SN2016coi is located right at the SVM boundary between the Ic-bl and Ib regions consistent with SN2016coi being similar to the SN Ib class.

\section{Summary \& Future Work}\label{sec:conc}
In this work, we have shown that PCA is a useful tool as a first step towards a data driven classification method for SESNe types. We used multi-class linear SVM's to explore different projections of SESNe spectra onto eigenspectra and found that the SESNe types are more distinguishable in the later phase ranges $t_{V_{max}}\approx10-15$ days relative to V-band maximum, rather than at peak light. We recommend that spectral follow-up of ZTF and LSST supernovae take these considerations into account. In addition, our classification method naturally provides a continuous, quantifiable method for characterizing ``transition'' SNe based on distance to class boundaries or centroids. We showed that our classification method identified both ``transition'' SNe and SNe with debated types previously identified in the literature, and we interpreted these SNe using our PCA eigenspectra and our SVM classification regions.

PCA is clearly a promising dimensionality reduction tool for SESNe, and there are many future projects that would use the work presented here as a starting point. In particular, the probability of a supernova's membership in one of the SESNe types could be calculated using the distance from its PCA projection to an SVM decision boundary. This provides a quantitative understanding of "transition" SNe like the type Ibc's, and should especially be explored as a function of phase.

\section{Acknowledgements}
We would like to thank Wolfgang Kerzendorf, Tyler Pritchard, David Hogg, Coryn Bailer-Jones, Markus Kromer, and Fritz Roepke for useful discussions. M.M. and the SNYU group are supported by the NSF CAREER award AST-1352405, by the NSF award AST-1413260 and by a Humboldt Faculty Fellowship.

This work is based (in part) on observations collected at the European Organisation for Astronomical Research in the Southern Hemisphere, Chile as part of PESSTO, (the Public ESO Spectroscopic Survey for Transient Objects Survey) ESO program 188.D-3003, 191.D-0935.

\end{document}